\documentclass[conference,10pt]{IEEEtran}
\usepackage{amssymb}
\usepackage{amsmath}
\usepackage{graphicx}
\usepackage{color}
\usepackage{multicol}
\usepackage{subfigure} 
\usepackage{bm}
\usepackage{latexsym}
\usepackage{stfloats}
\usepackage{float}
\usepackage{exscale}
\usepackage{relsize}
\usepackage{cite}
\usepackage{cases}
\usepackage{algorithm}
\usepackage{algpseudocode}
\usepackage{amsthm}
\usepackage{epsfig} 
\usepackage{epstopdf}
\usepackage{breqn}
\usepackage{enumerate}
\usepackage{multirow}
\usepackage[utf8]{inputenc}
\usepackage{algorithmicx}
\usepackage{makecell} 

\begin{document}
\title{\huge Deep Learning-Based Rate-Adaptive CSI Feedback for Wideband XL-MIMO Systems in the Near-Field Domain} 
\vspace{-5ex}
\author{ Zhenyu Liu, Yi Ma,  and Rahim Tafazolli\\
	{6GIC, Institute for Communication Systems, University of Surrey, Guildford, UK, GU2 7XH}\\
	{ Emails: (zhenyu.liu, y.ma, r.tafazolli)@surrey.ac.uk }
}
\maketitle

\begin{abstract}
Accurate and efficient channel state information (CSI) feedback is crucial for unlocking the substantial spectral efficiency gains of extremely large-scale MIMO (XL-MIMO) systems in future 6G networks. However, the combination of near-field spherical wave propagation and frequency-dependent beam split effects in wideband scenarios poses significant challenges for CSI representation and compression. This paper proposes \emph{WideNLNet-CA}, a rate-adaptive deep learning framework designed to enable efficient CSI feedback in wideband near-field XL-MIMO systems. WideNLNet-CA introduces a lightweight encoder–decoder architecture with multi-stage downsampling and upsampling, incorporating computationally efficient residual blocks to capture complex multi-scale channel features with reduced overhead. A novel compression ratio adaptive module with feature importance estimation is introduced to dynamically modulate feature selection based on target compression ratios, enabling flexible adaptation across a wide range of feedback rates using a single model. Evaluation results demonstrate that WideNLNet-CA consistently outperforms existing compressive sensing and deep learning-based works across various compression ratios and bandwidths, while maintaining fast inference and low model storage requirements.
\end{abstract}


\section{Introduction}
Extremely large-scale multiple-input multiple-output (XL-MIMO) is a key enabler for enhancing spectral efficiency and throughput in future 6G wireless networks~\cite{1}. By deploying a vast number of antennas, XL-MIMO can achieve significant beamforming gains and offer up to a tenfold improvement in spectral efficiency over conventional massive MIMO systems~\cite{channel_model}, thereby supporting ultra-high data rates. However, in frequency division duplexing (FDD) systems, achieving these gains requires accurate channel state information (CSI) at the base station (BS), obtained via feedback from user equipment (UE). Due to the large number of antennas and subcarriers, the resulting CSI matrices in XL-MIMO systems are extremely high-dimensional, placing a substantial burden on uplink bandwidth. Efficient CSI compressed feedback mechanisms are thus essential for the practical deployment of XL-MIMO.

To address the overhead of CSI acquisition in XL-MIMO systems, compressive sensing (CS) techniques have been extensively explored. By leveraging the potential sparsity of XL-MIMO channels in the angular-delay domain, CS-based methods~\cite{cs_xlmimo1, BSPD} achieve higher compression efficiency than conventional codebook-based approaches~\cite{codebook_xlmimo2}. However, the performance of CS techniques depends heavily on strict channel sparsity assumptions, which may not always hold. Additionally, CS algorithms often involve iterative and computationally intensive recovery procedures, introducing latency.

More recently, deep learning (DL)-based CSI feedback methods~\cite{cui2022, full_wang2023, bifull_2024lin} have demonstrated superior performance in both compression efficiency and reconstruction accuracy. These methods outperform traditional techniques by effectively leveraging channel characteristics such as spatial and spectral correlations \cite{cui2022, full_wang2023} and bi-directional correlations \cite{bifull_2024lin}. The inclusion of DL-based CSI feedback as a key use case in 3GPP TR 38.843~\cite{3gpp2023csi} further highlights its promise. Nevertheless, existing DL-based solutions—mainly developed for far-field massive MIMO—face significant challenges in near-field XL-MIMO systems. First, near-field propagation exhibits spherical rather than planar wavefronts, resulting in more complex CSI structures. Second, the enlarged antenna arrays and increased subcarrier counts substantially inflate the CSI dimensionality, leading to severe memory and computational burdens without proper optimization.

To cope with these near-field challenges, a narrowband DL-based framework, ExtendNLNet~\cite{10720839}, was proposed. By incorporating a non-local block to extract spatially extended CSI features, ExtendNLNet improves reconstruction accuracy over methods designed for the far field. However, in wideband XL-MIMO systems, the beam split effect~\cite{split_eff}—where spatial channel directions vary with frequency—introduces frequency-dependent sparsity in the angle domain, significantly impairing performance. Moreover, ExtendNLNet and similar methods use fixed compression ratios, lacking adaptability to dynamic bandwidth constraints and resulting in high storage demands. Although CsiNet+~\cite{8972904} enables variable-rate feedback for wideband massive MIMO, it requires over one billion parameters in the compression module in XL-MIMO systems and still relies on separate decoder networks for each rate.

In this paper, we propose \textit{WideNLNet-CA}, a novel rate-adaptive deep learning framework designed for wideband near-field XL-MIMO CSI feedback, accounting for the challenges posed by spherical wave propagation, the beam split effect and high computational complexity. To the best of our knowledge, \textit{WideNLNet-CA} is the first DL-based approach developed for CSI feedback in wideband near-field XL-MIMO systems.

The main contributions are summarized as follows:
\begin{itemize}
\item We propose \textit{WideNLNet-CA}, a DL-based framework designed for efficient and robust CSI feedback in wideband near-field XL-MIMO systems. Its hierarchical encoder-decoder architecture employs multi-stage downsampling/upsampling and lightweight residual blocks to learn complex, multi-scale features of the wideband near-field channel while maintaining high recovery accuracy.
\item We develop a novel compression ratio adaptive module that uses a target rate-conditioned mechanism and feature importance estimation to dynamically modulate and select salient channel features. This enables the network to operate over a wide range of compression ratios with a single trained model, efficiently reducing storage and training overhead.
\item Simulation results show that WideNLNet-CA consistently and significantly outperforms state-of-the-art CS- and DL-based CSI feedback methods across diverse compression ratios and system bandwidths with fast inference time and low storage cost. 
\end{itemize}

\section{System Model}
\label{sec:system_model}

\subsection{Signal Model}

We consider a downlink wideband XL-MIMO system operating in the near-field and employing orthogonal frequency-division multiplexing (OFDM). The BS is equipped with a uniform linear array (ULA) of $N$ antennas, while the UE is equipped with a single antenna. The system operates over $M$ subcarriers with carrier frequency $f_c$ and bandwidth $B$.

The received signal at the $m$-th subcarrier is given by
\begin{align}
y_m = \mathbf{h}_m^H \mathbf{v}_m x_m + n_m,
\end{align}
where $\mathbf{h}_m \in \mathbb{C}^{N \times 1}$ denotes the channel vector, $\mathbf{v}_m \in \mathbb{C}^{N \times 1}$ is the precoding vector, $x_m$ is the transmitted symbol, $n_m$ is the noise, and $(\cdot)^H$ denotes the Hermitian transpose.

\subsection{Near-Field Wideband Channel Model}
\label{subsec:channel_model}
For the $m$-th subcarrier ($m \in \{0,1,\ldots,M-1\}$), we adopt the spherical wave propagation model to capture near-field effects. The channel vector is modeled as
\begin{align}
\mathbf{h}_m = \sqrt{\frac{N}{L}} \sum_{l=0}^{L-1} g_{l,m} e^{-j\frac{2\pi}{\lambda_m} r_l} \mathbf{a}(\vartheta_l, r_l, f_m),
\end{align}
where $g_{l,m}$ is the complex gain of the $l$-th path at subcarrier $m$, and $L$ is the total number of propagation paths. The frequency of the $m$-th subcarrier is
\begin{align}
f_m = f_c + \frac{2m - M}{2M} B,
\end{align}
with $\lambda_m = c / f_m$ representing the corresponding wavelength, where $c$ is the speed of light.

The array response vector $\mathbf{a}(\vartheta_l, r_l, f_m)$ characterizes the spherical wavefront between the scatter and the BS array, which is modeled as
\begin{align}
\mathbf{a}(\vartheta_l, r_l, f_m) = \frac{1}{\sqrt{N}}[e^{j\phi_{l,m}^{(0)}}, e^{j\phi_{l,m}^{(1)}}, \ldots, e^{j\phi_{l,m}^{(N-1)}}]^T,
\end{align}
where $\phi_{l,m}^{(n)} = -\frac{2\pi}{\lambda_m}(r_l^{(n)} - r_l)$. Here, $r_l^{(n)} = \sqrt{r_l^2 + \delta_n^2d^2 - 2r_l\delta_nd\sin\vartheta_l}$ represents the distance from the scatter to the $n$-th BS antenna \cite{channel_model}, with $\delta_n = n - \frac{N-1}{2}$ for $n \in \{0, 1, \ldots, N-1\}$. Additionally, $d = \frac{\lambda_c}{2}$ is the antenna spacing and $\lambda_c = \frac{c}{f_c}$ is the carrier wavelength.

This near-field channel model reveals two key characteristics of wideband XL-MIMO that are not fully accounted for by existing CSI feedback works~\cite{10720839,BSPD,mourya2023}:

1) \textbf{Near-field propagation:} Unlike conventional massive MIMO models that approximate $\phi_{l,m}^{(n)}$ as linear in $\delta_n d \sin\vartheta_l$ under the far-field assumption, XL-MIMO operates at distances comparable to the Fraunhofer distance (e.g., $2D^2/\lambda_c$, where $D$ is the array aperture), necessitating a spherical model that accounts for distance variation across the array elements.

2) \textbf{Beam split effect:} The array response vector becomes frequency-dependent when $f_m \neq f_c$, leading to spatial dispersion of energy across subcarriers. This beam split phenomenon \cite{NFBF_Cui2023} invalidates the strong sparsity assumptions in the angular-delay domain \cite{BSPD}, resulting in degraded feedback performance.

\subsection{CSI Feedback Framework}

The downlink CSI matrix in the spatial-frequency domain is denoted as $\tilde{\mathbf{H}} = [\mathbf{h}_1, \ldots, \mathbf{h}_M]^H \in \mathbb{C}^{M \times N}$. To exploit channel sparsity, we transform $\tilde{\mathbf{H}}$ into the angular-delay domain using a 2D discrete Fourier transform (DFT):
\begin{align}
\mathbf{H} = \mathbf{F}_d^H \tilde{\mathbf{H}} \mathbf{F}_a \in \mathbb{C}^{M \times N},
\end{align}
where $\mathbf{F}_d$ and $\mathbf{F}_a$ are $M \times M$ and $N \times N$ unitary DFT matrices, respectively.

To accommodate real-valued neural networks, we split the complex matrix $\mathbf{H}$ into its real and imaginary parts, transpose it, and normalize both parts to $[0, 1]$. This results in an input tensor $\mathbf{H}_{\text{ts}} \in \mathbb{R}^{N \times M \times 2}$.

We define the compression ratio (CR) as
\begin{align}
\beta = \frac{K_{\text{code}}}{2NM},
\end{align}
where $K_{\text{code}}$ is the dimensionality of the compressed codeword. In conventional feedback systems, one must train separate encoder–decoder pairs for each CR:
\begin{align}
\mathbf{s} &= f_{\text{en}}(\mathbf{H}_{\text{ts}}; \Phi_\beta), \\
\hat{\mathbf{H}}_{\text{ts}} &= f_{\text{de}}(\mathbf{s}; \Psi_\beta),
\end{align}
with $\Phi_\beta$ and $\Psi_\beta$ being parameters specific to CR = $\beta$.

For wideband XL-MIMO, training and storing multiple models for different CRs is computationally inefficient and requires significant storage. To overcome this, we propose a rate-adaptive feedback framework where the CR $\beta$ is treated as a conditional input to a shared encoder–decoder model:
\begin{align}
\mathbf{s} &= f_{\text{en}}(\mathbf{H}_{\text{ts}}, \beta; \Phi), \\
\hat{\mathbf{H}}_{\text{ts}} &= f_{\text{de}}(\mathbf{s}, \beta; \Psi),
\end{align}
where $\Phi$ and $\Psi$ are shared parameters across all CRs. This approach enables efficient CSI compression and reconstruction across variable feedback rates using a single model.

\section{Proposed WideNLNet-CA Architecture}
\label{sec:proposed_architecture}

This section introduces the architecture of the proposed rate-adaptive CSI feedback network, WideNLNet-CA, designed for wideband XL-MIMO systems in the near-field domain.

\begin{figure*}[thpb]
    \centering
    \includegraphics[width=0.76\linewidth]{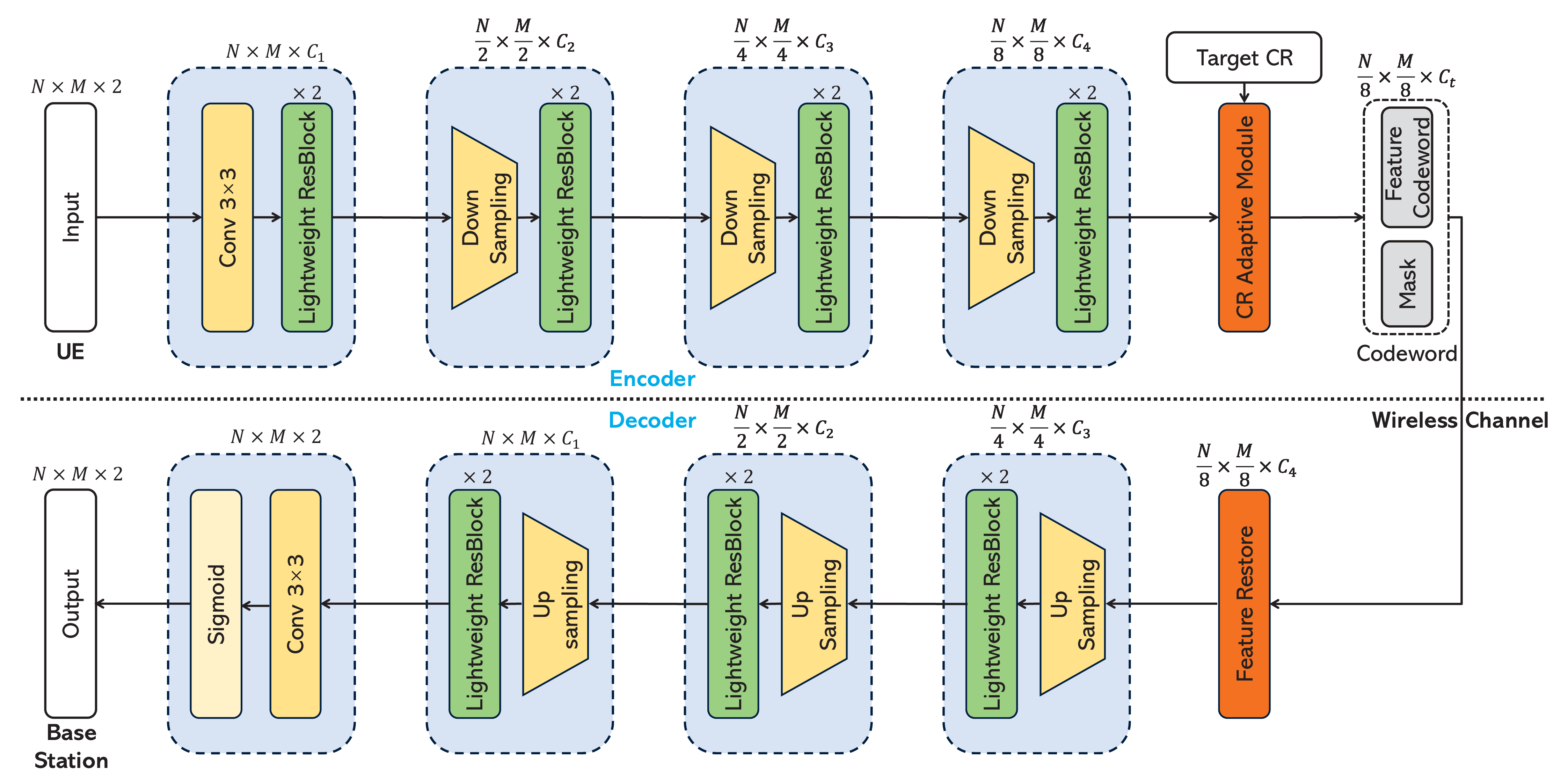}
    \vspace*{-6mm}
    \caption{Overall architecture of the proposed WideNLNet-CA.}
    \vspace*{-6mm}
    \label{fig:overall_architecture}
\end{figure*}

\subsection{Encoder Network}
\label{ssec:encoder_details}

\begin{figure}[!t]
    \centering
    \includegraphics[width=0.76\linewidth]{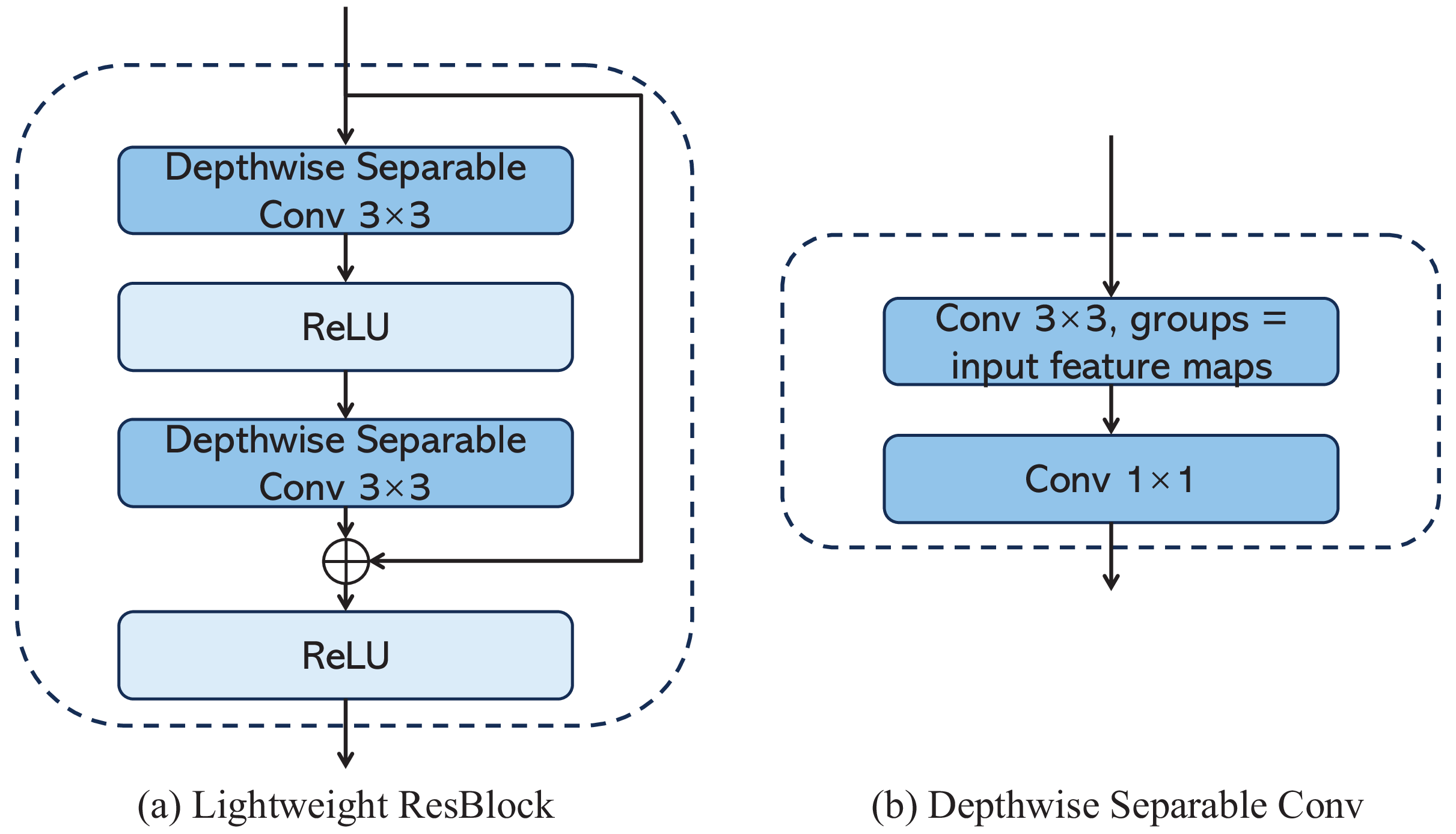}
    \vspace*{-3mm}
    \caption{Structure of Lightweight Residual Block based on depthwise separable convolutions. }
    \vspace*{-6mm}
    \label{fig:resblock_details}
\end{figure}

The encoder, illustrated in the upper path of Fig.~\ref{fig:overall_architecture}, transforms the high-dimensional input CSI tensor $\mathbf{H}_{\text{ts}} \in \mathbb{R}^{ N \times M \times 2}$ into a compact latent representation.

The encoder begins with a $3 \times 3$ convolutional layer that extracts local angular-delay features from $\mathbf{H}_{\text{ts}}$ and changes the channel dimension from $2$ to $C_1$. This is followed by a Lightweight Residual Block (LightResBlock), whose structure is shown in Fig.~\ref{fig:resblock_details}(a). Each LightResBlock utilizes depthwise separable convolutions \cite{Li_2022_CVPR} to significantly reduce computational cost and parameter count, making them suitable for large CSI matrices. Residual connections further facilitate deep network training by mitigating gradient vanishing. After this initial stage, the feature map has dimensions $(N, M, C_1)$.

The encoder then applies a sequence of three downsampling stages, each consisting of:
\begin{itemize}
    \item \textbf{Downsampling Layer}: A strided $3 \times 3$ convolution layer that reduces both angular and delay dimensions by half, i.e., $N \to N/2$, $M \to M/2$.
    \item \textbf{LightResBlock}:A subsequent LightResBlock that extracts and refines features at the new resolution.
\end{itemize}

This hierarchical design enables the encoder to extract increasingly abstract features, transitioning from fine-grained local patterns in early layers to high-level representations in deeper layers. The sizes of feature maps are reduced stepwise from $(N, M)$ to $(N/8, M/8)$, while the number of feature maps increases, eventually reaching $C_4$. The final output of this part of the encoder, with shape $(N/8, M/8, C_4)$, is forwarded to the CR Adaptive Module.

For the proposed non-adaptive WideNLNet (WideNLNet without CAM), the encoder structure is similar but concludes by directly producing $C_t$ feature maps (i.e., via an additional $3 \times 3$ convolution adjusting channels from $C_4$ to $C_t = 128\beta$), specific to a fixed CR $\beta$. This variant requires training a separate model for each desired CR.

\vspace*{-1mm}

\subsection{CR Adaptive Module}
\label{ssec:cam_module}
\vspace*{-1mm}

\begin{figure*}[!t]
    \centering
    \includegraphics[width=0.72\linewidth]{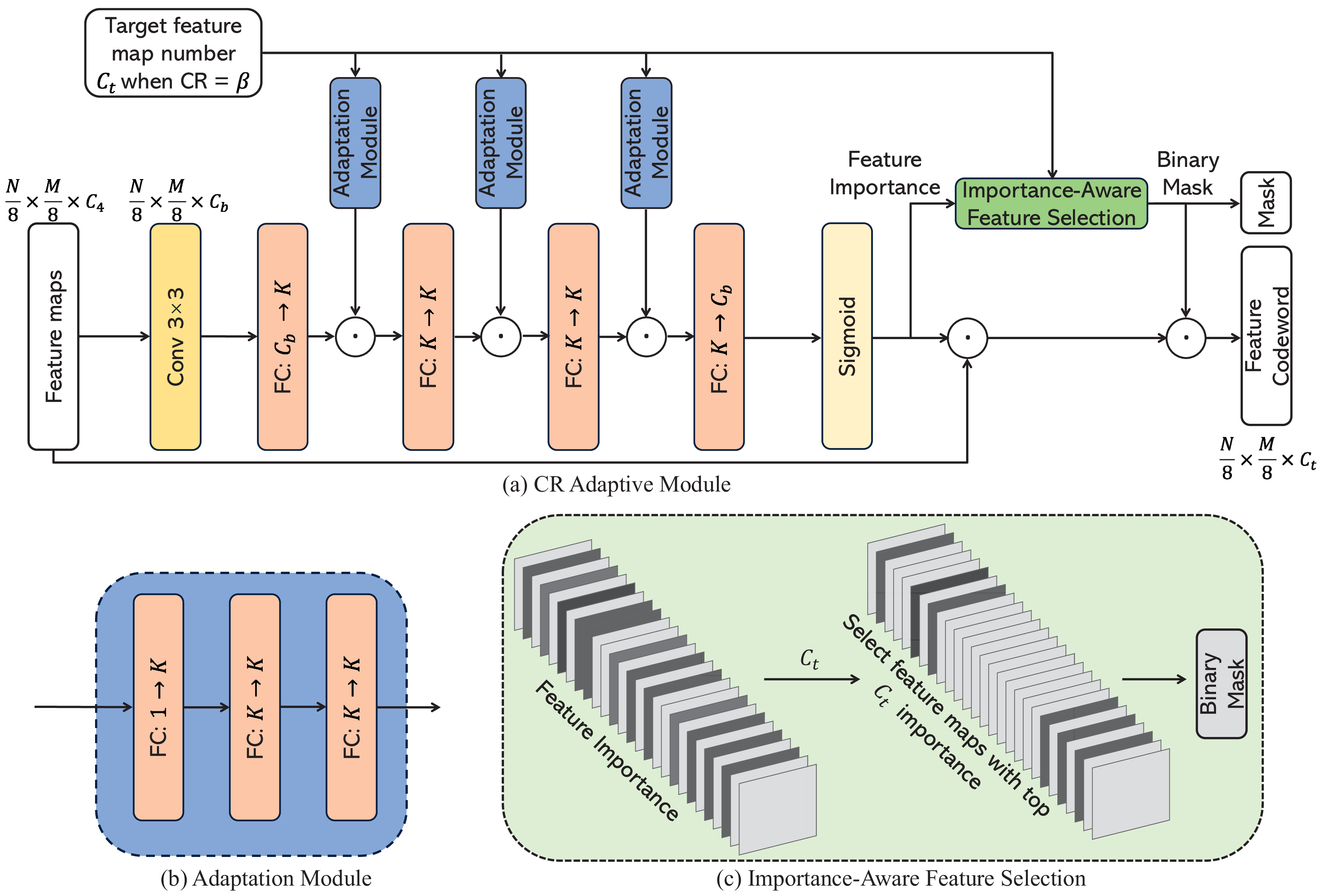} 
    \vspace*{-3mm}
    \caption{Architecture of the CR Adaptive Module (CAM). }
    \vspace*{-6mm}
    \label{fig:CAM_details}
\end{figure*}

To support dynamic adaptation to various target CSI feedback rates without requiring retraining or maintaining multiple models, we introduce the CR Adaptive Module (CAM), as illustrated in Fig.~\ref{fig:CAM_details}. The CAM intelligently processes and selects features based on a target feature map count, denoted as $C_t$, which corresponds to the desired CR $\beta$. 

The CAM receives feature maps from the encoder, which first undergo an initial $3 \times 3$ convolution. This step refines the input features and adjusts their channel dimension to an internal working depth $C_b$, producing an intermediate feature map $\mathbf{z}_{\text{conv}}$, which can be the maximum feature maps required in all desired CRs.

A crucial CAM function is generating a Feature Importance representation dynamically adapting to $C_t$. This is derived from $\mathbf{z}_{\text{conv}}$ by processing its $C_b$-dimensional feature vector at each spatial location through a specialized path (Fig.~\ref{fig:CAM_details}(a)). This path consists of four Fully Connected (FC) layers, transforming features through intermediate $K$-dimensional representations and back to $C_b$ dimensions, interspersed with three Adaptation Modules. Each Adaptation Module (Fig.~\ref{fig:CAM_details}(b)), a three-layered FC network conditioned on $C_t$, generates a modulation vector $\mathbf{m}_j$ that adaptively scales the output of the preceding FC layer. This cascaded modulation allows multi-stage, rate-dependent feature adjustment. The transformation of $C_t$ to $\mathbf{m}_j$ within an Adaptation Module is achieved by:
\begin{subequations}
    \begin{align}
        \mathbf{m}^{(1)}_j &= \text{ReLU}(\mathbf{W}^{(1)}_{\text{adapt}} \cdot C_t + \mathbf{b}^{(1)}_{\text{adapt}}), \\
        \mathbf{m}^{(2)}_j &= \text{ReLU}(\mathbf{W}^{(2)}_{\text{adapt}} \cdot \mathbf{m}^{(1)}_j + \mathbf{b}^{(2)}_{\text{adapt}}), \\
        \mathbf{m}_j &= \text{Sigmoid}(\mathbf{W}^{(3)}_{\text{adapt}} \cdot \mathbf{m}^{(2)}_j + \mathbf{b}^{(3)}_{\text{adapt}}),
    \end{align}
\end{subequations}
where $\mathbf{W}^{(k)}_{\text{adapt}}$ and $\mathbf{b}^{(k)}_{\text{adapt}}$ are learnable parameters.

The previously generated importance tensor, with dimensions $(N/8 \times M/8 \times C_b)$, is then transformed into a spatially-aware feature importance map $\bm{\alpha}$ by applying a Sigmoid activation function. This map $\bm{\alpha}$ contains normalized importance scores for each feature at every spatial location. Subsequently, the intermediate feature map $\mathbf{z}_{\text{conv}}$ is modulated by $\bm{\alpha}$ via element-wise multiplication to emphasize salient features:
\begin{equation}
    \mathbf{z}_{\text{modulated}} = \mathbf{z}_{\text{conv}} \odot \bm{\alpha}.
\end{equation}

To achieve the precise target feature map count $C_t$, an ``Importance-Aware Feature Selection" mechanism is then employed, as depicted in Fig.~\ref{fig:CAM_details}(c). This involves first deriving global channel-wise importance scores by spatially aggregating the per-location scores within $\bm{\alpha}$ for each of the $C_b$ channels through averaging. These $C_b$ aggregated scores are ranked, and the top $C_t$ channels are selected. Based on this selection, a binary channel mask $\mathbf{M}$ is generated, which has ones for the chosen $C_t$ channels and zeros for the remainder. This mask is applied to the modulated features $\mathbf{z}_{\text{modulated}}$ to produce the final rate-adapted output:
\begin{equation}
    \mathbf{z}_{\text{CAM\_out}} = \mathbf{z}_{\text{modulated}} \odot \mathbf{M}.
\end{equation}
The resulting $\mathbf{z}_{\text{CAM\_out}}$ thus effectively contains $C_t$ active channels, matching the target rate.

The CAM's output, $\mathbf{z}_{\text{CAM\_out}}$, termed the  ``Feature Codeword" in Fig.~\ref{fig:overall_architecture}, is then transmitted. Notably, the overhead for transmitting the mask information $\mathbf{M}$ (i.e., $C_b$ bits for a simple bitmask representation) is negligible in wideband XL-MIMO systems compared to the feature codeword payload.

\subsection{Decoder Network}
\label{ssec:decoder_details}

The decoder reconstructs the high-dimensional CSI tensor from the received feature codeword $\hat{\mathbf{z}}_{\text{CAM\_out}} \in \mathbb{R}^{(N/8) \times (M/8) \times C_t}$ (only active feature maps are received and others padded with zeros to form $C_b$ channels). The decoder structure mirrors the encoder, employing a sequence of upsampling layers and LightResBlocks.

The reconstruction begins with a feature restore block (i.e., a $1 \times 1$ convolution) to adjust the input channels from $C_t$ to $C_4$. The following three upsampling stages each consist of:
\begin{itemize}
    \item \textbf{Upsampling Layer}: A transposed $3 \times 3$ convolution layer doubles angular and delay dimensions.
    \item \textbf{LightResBlock}: Refines features at the new resolution.
\end{itemize}

After the final upsampling stage and LightResBlock (outputting $C_1$ channels), a $3 \times 3$ convolution adjusts the channel dimension to $2$, and a Sigmoid activation function produces the reconstructed CSI tensor $\hat{\mathbf{H}}_{\text{ts}} \in \mathbb{R}^{N \times M \times 2}$, normalized to the $[0,1]$ range.

The symmetric encoder-decoder structure, combined with lightweight and efficient modules, enables the model to learn compact yet expressive representations of complex near-field XL-MIMO channels and achieve high-fidelity reconstruction under diverse compression constraints.

\subsection{Training Process}
\label{ssec:training_process}

WideNLNet-CA is trained end-to-end by minimizing the mean squared error (MSE) between the original CSI tensor $\mathbf{H}_{\text{ts}}$ and the reconstructed tensor $\hat{\mathbf{H}}_{\text{ts}}$:
\begin{equation}
    \mathcal{L} = \frac{1}{D_{\text{size}}} \sum_{i=1}^{D_{\text{size}}} \| \mathbf{H}_{\text{ts}}^{(i)} - \hat{\mathbf{H}}_{\text{ts}}^{(i)} \|_F^2,
    \label{eq:loss_mse}
\end{equation}
where $D_{\text{size}}$ is the size of the training dataset, $\mathbf{H}_{\text{ts}}^{(i)}$ is the $i$-th input sample, $\hat{\mathbf{H}}_{\text{ts}}^{(i)}$ is its reconstruction, and $\|\cdot\|_F^2$ denotes the squared Frobenius norm.

To achieve rate-adaptiveness, a multi-rate training strategy is adopted. For each training iteration or batch, a target compression ratio $\text{CR}_{\text{target}}$ is randomly selected from a predefined set (e.g., $\{1/4, 1/16, 1/64\}$), which determines the corresponding $C_t$. The CAM is then configured with this $C_t$. This allows a single model to operate effectively across multiple CRs without the need for separate networks.

\vspace*{-3mm}

\section{Performance Evaluation}
\label{sec:performance_evaluation}
\vspace*{-1mm}
\subsection{Simulation Setup}
\label{ssec:sim_setup}

 We consider a wideband near-field XL-MIMO system in which the BS is equipped with a ULA of $N = 256$ antennas. The system operates over $M = 256$ subcarriers centered at a carrier frequency of $f_c = 100$~GHz with a total bandwidth of $B = 10$~GHz. Near-field channels are generated based on the spherical wave propagation model described in Section~\ref{subsec:channel_model}, with parameters aligned with those in~\cite{channel_model}.

A dataset comprising 45{,}000 independent channel realizations is used, with 35{,}000 samples for training, 5{,}000 for validation, and 5{,}000 for testing, following the setup in~\cite{10720839}. The network is implemented in PyTorch and trained using the Adam optimizer with a learning rate of $3 \times 10^{-4}$ for 200 epochs. The encoder and decoder use the number of feature maps $C_1 = 64$, $C_2 = 64$, $C_3 = 128$, and $C_4 = 128$. The CAM is configured with an internal depth $C_b = 32$ and a latent dimension $K = 64$ in its feature selection path.

The following methods are selected for comparison:
\begin{itemize}
    \item \textbf{ExtendNLNet~\cite{10720839}:} A DL-based CSI feedback model originally designed for narrowband XL-MIMO. For fair comparison under wideband conditions, its input dimensions are adjusted, and lightweight optimizations are applied~\cite{lightweight1} to control model size and complexity.
    \item \textbf{BSPD (Beam Split Pattern Detection)~\cite{BSPD}:} A CS-based approach that exploits angular-delay sparsity and beam split pattern mapping for wideband CSI recovery.
\end{itemize}

The following two evaluation metrics are adopted, in line with prior studies~\cite{10720839, cui2022, full_wang2023}.  First, the normalized mean squared error (NMSE) quantifies the reconstruction accuracy of the channel matrix and is defined as
\begin{equation}
    \text{NMSE (dB)} = 10\log_{10}\left(\mathbb{E}\left\{\frac{\|\mathbf{H} - \hat{\mathbf{H}}\|_F^2}{\|\mathbf{H}\|_F^2}\right\}\right),
\end{equation}
where $\mathbf{H}$ and $\hat{\mathbf{H}}$ denote the original and reconstructed complex CSI matrices, and $\mathbb{E}\{\cdot\}$ represents the expectation over the test dataset.
Second, the cosine similarity, denoted by $\rho$, is expressed as
\begin{equation}
    \rho = \mathbb{E}\left\{\frac{1}{M}\sum_{m=1}^{M}\frac{|\hat{\mathbf{h}}_m^H \mathbf{h}_m|}{\|\hat{\mathbf{h}}_m\|_2 \|\mathbf{h}_m\|_2}\right\},
\end{equation}
where $\mathbf{h}_m$ and $\hat{\mathbf{h}}_m$ represent the ground truth and reconstructed CSI vectors at the $m$-th subcarrier.

\begin{figure}[t]
    \centering 
    \subfigure[NMSE]{
        \includegraphics[width=0.39\textwidth]{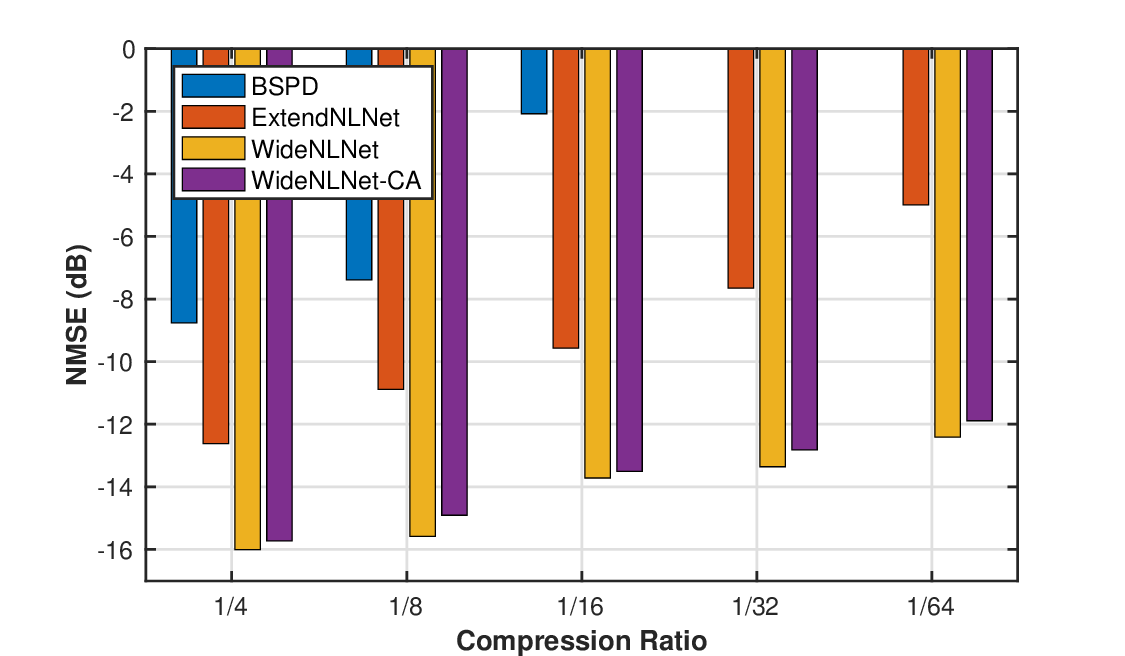} 
        \label{fig:cr_nmse}
    } \\ \vspace{-11pt}
    \subfigure[Cosine Similarity]{
        \includegraphics[width=0.39\textwidth]{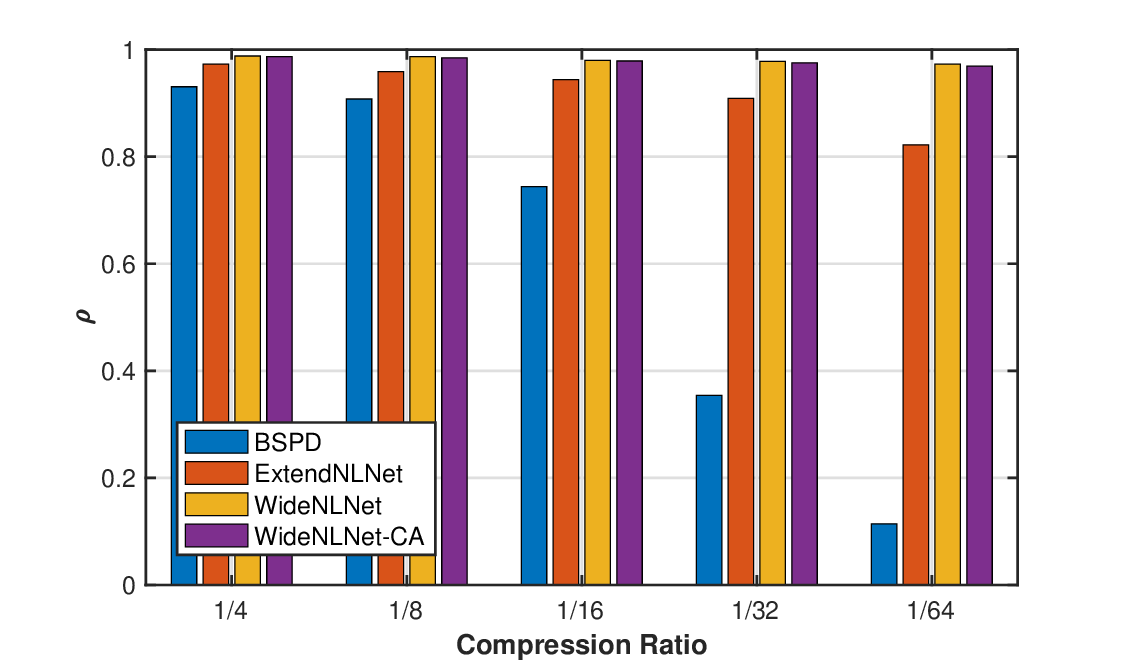}
        \label{fig:cr_rho}
    } 
    \vspace{-6pt}
    \caption{Performance comparison across different compression ratios.}
    \vspace*{-6mm}
    \label{fig:cr_comparisons}
\end{figure}

\subsection{Performance Across Different Compression Ratios}
\label{ssec:perf_crs}

Fig.~\ref{fig:cr_comparisons} compares the performance of WideNLNet-CA with that of its non-adaptive counterpart (WideNLNet), ExtendNLNet, and BSPD under varying CRs ranging from $1/4$ to $1/64$.
As illustrated in Fig.~\ref{fig:cr_comparisons}, WideNLNet-CA consistently outperforms all baseline methods across the full range of CRs. The performance gap is particularly significant at high compression levels (e.g., CR = $1/32$ and $1/64$). At CR = $1/64$, WideNLNet-CA achieves an NMSE of approximately $-11.8$~dB, outperforming ExtendNLNet ($-5.0$~dB) and BSPD, which fails to preserve acceptable accuracy. In terms of cosine similarity, WideNLNet-CA also maintains strong alignment with the original channels, indicating robust channel reconstruction quality even at extremely low feedback rates.

These results validate the effectiveness of the proposed rate-adaptive CAM design. Although the base WideNLNet model achieves strong performance when trained separately for each compression ratio, WideNLNet-CA offers a critical advantage by delivering comparable accuracy while supporting multiple rates with a single model, eliminating the need for training and deploying multiple networks. BSPD performs reasonably well at moderate CRs but degrades sharply under high compression due to its reliance on angular-delay sparsity, which is not guaranteed in wideband near-field scenarios.

\subsection{Performance Across Varying Bandwidths}
\label{ssec:perf_bandwidth}

\begin{figure}[!t]
    \centering
    \subfigure[NMSE]{
        \includegraphics[width=0.38\textwidth]{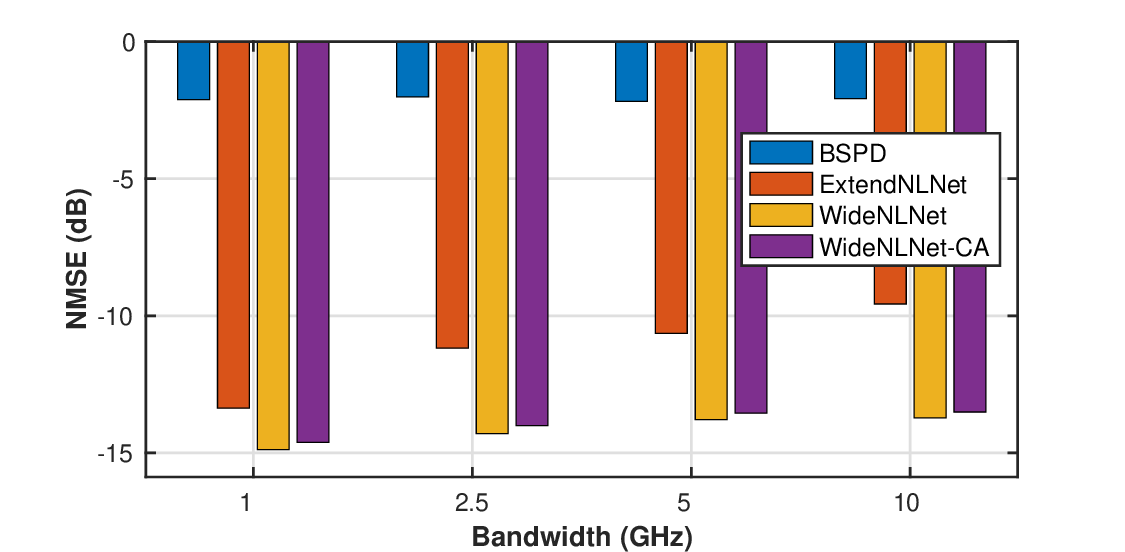}
        \label{fig:bw_nmse}
    } \\ \vspace*{-3.9mm}
    \subfigure[Cosine Similarity]{
        \includegraphics[width=0.38\textwidth]{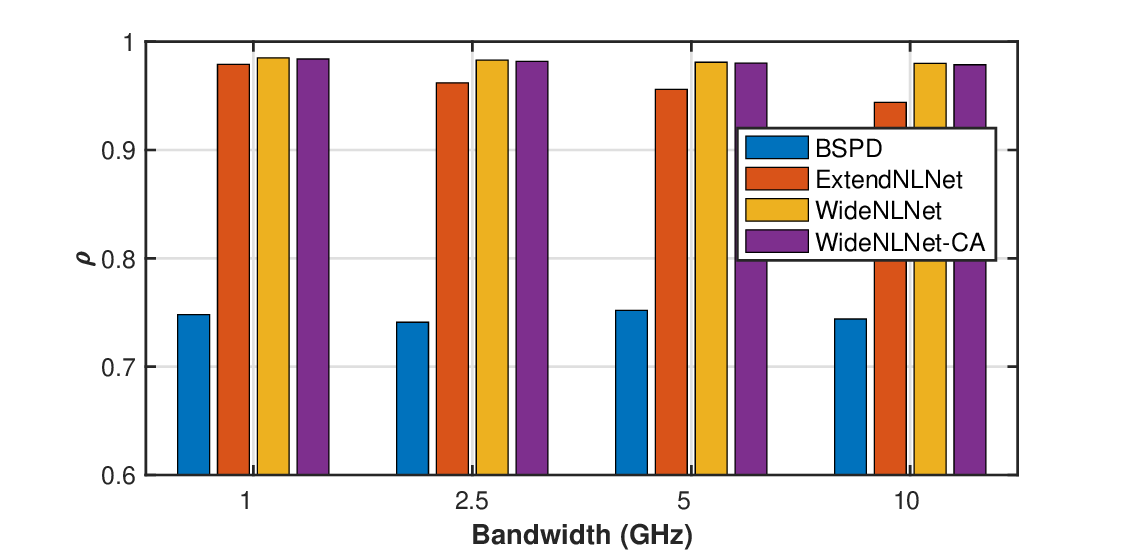}
        \label{fig:bw_rho}
    }
    \vspace*{-3mm}
    \caption{Performance under varying system bandwidths when CR $= 1/16$.}
    \vspace*{-5mm}
    \label{fig:bw_comparisons}
\end{figure}

To evaluate robustness against beam split effects, we vary the system bandwidth from 1~GHz to 10~GHz while fixing the compression ratio at $1/16$.  As shown in Fig.~\ref{fig:bw_comparisons}, increasing bandwidth intensifies the beam split effect, thereby complicating CSI feedback. ExtendNLNet suffers substantial degradation as its design is optimized for narrowband settings and lacks mechanisms to account for frequency-dependent spatial shifts. BSPD demonstrates relative robustness to bandwidth variation due to its beam split modeling, but its overall performance remains limited under near-field conditions.

In contrast, both WideNLNet and WideNLNet-CA maintain strong performance across the entire bandwidth range. This demonstrates their ability to learn generalized channel representations that capture near-field propagation and frequency dispersion. Although slight degradation is observed at 10~GHz, the loss is significantly smaller than that of baseline methods, affirming the proposed architecture’s robustness.
\vspace*{-1mm}
\subsection{Computational Complexity}
\label{ssec:complexity}
\vspace*{-1mm}
Table~\ref{tab:complexity_comparison} compares the computational complexity of the evaluated methods in terms of model size, inference time, and the number of models required to support $N_{\text{CR}}$ distinct CRs. Inference latency is measured on an NVIDIA RTX 4090 GPU and Intel Xeon w7-3465X CPU under CR = $1/16$.

\begin{table}[!t]
  \centering
  \caption{Computational Complexity Comparison When CR = $1/16$ (ENL:
ExtendNLNet, WNL: WideNLNet, M: Million)}
\vspace*{-1mm}
  \begin{tabular}{|c|c|c|c|c|}
    \hline
    \textbf{Metric} & \textbf{BSPD} & \textbf{ENL} & \textbf{WNL} & \textbf{WNL-CA} \\
    \hline
    Parameters & $1.07$~M & $8.39$~M & $1.47$~M & $1.52$~M \\
    \hline
    Inference Time & $130$~ms & $2.51$~ms & $1.57$~ms & $1.64$~ms \\
    \hline
    \makecell{Models for $N_{\text{CR}}$ CRs} & $1$ & $N_{\text{CR}}$ & $N_{\text{CR}}$ & $1$ \\
    \hline
  \end{tabular}
  \vspace*{-6mm}
  \label{tab:complexity_comparison}
\end{table}

Although BSPD uses fewer parameters, its inference time is significantly higher due to CPU-bound iterative recovery. ExtendNLNet exhibits the highest model complexity, even after lightweight refinement. WideNLNet and WideNLNet-CA are both highly efficient, with inference times below 1.7~ms and compact model sizes. Importantly, WideNLNet-CA supports multiple CRs using a single trained model, substantially reducing storage and deployment costs. The minor increase in parameter count compared to WideNLNet is well-justified by the significant gains in flexibility and performance.

\vspace*{-1mm}
\section{Conclusion}

This paper presented WideNLNet-CA, a rate-adaptive deep learning framework for CSI feedback in wideband near-field XL-MIMO systems. By integrating a lightweight encoder–decoder architecture with a novel CR adaptive module, the proposed network enables accurate CSI reconstruction across diverse compression ratios using a single unified model. Simulation results demonstrated that WideNLNet-CA consistently outperforms existing compressive sensing and deep learning-based methods under various feedback rates and system bandwidths, particularly in challenging scenarios involving severe compression and pronounced beam split effects. Furthermore, its reduced storage requirement and real-time inference capability make WideNLNet-CA an efficient solution for practical deployment in future XL-MIMO systems.

\section*{Acknowledgement}
This work was partially funded by the UKRI-EPSRC under Intelligent Spectrum Innovation (ICON) programme (APP55159).

\bibliographystyle{IEEEtran}
\bibliography{Refer} 

\end{document}